\documentclass[twocolumn,prl,aps,showpacs,superscriptaddress]{revtex4-1}
\usepackage{graphicx,graphics}
\usepackage{amsmath}
\usepackage{amssymb}
\usepackage{epstopdf}
\usepackage{amstext}
\usepackage{amsthm}
\usepackage{amsfonts}
\usepackage{latexsym}
\usepackage{array}
\usepackage{bm}
\usepackage{xfrac}
\usepackage{lipsum}
\usepackage[toc,page]{appendix}
\usepackage{subfigure}
\newcommand{\igbjd}[1]{}\newcommand{\beqa}{\begin{eqnarray}}
\newcommand{\eeqa}{\end{eqnarray}}
\newcommand{\beq}{\begin{equation}}
\newcommand{\eeq}{\end{equation}}
\usepackage[normalem]{ulem}
\usepackage[colorlinks=true,citecolor=Cerulean,linkcolor=RubineRed,urlcolor=Cerulean,pdftex]{hyperref}
\usepackage{xcolor}

\usepackage[free-standing-units=true]{siunitx}

\definecolor{Cerulean}{rgb}{0.,0.59,0.835}
\definecolor{RubineRed}{rgb}{0.61,0.07,0.12}

\newcommand{\1}{{\text A}}

\begin{document}
\title{Self-Assembled Chains and Solids of Dipolar Atoms in a Multilayer}
\author{G. Guijarro}
\affiliation{Departament de F\'isica, Campus Nord B4-B5, 
Universitat Polit\`ecnica de Catalunya,
E-08034 Barcelona, Spain}
\affiliation{Theoretische Physik, Universit\"at des Saarlandes, Campus E2.6, D-66123 Saarbr\"ucken, Germany}
\author{G. E. Astrakharchik}
\affiliation{Departament de F\'isica, Campus Nord B4-B5, Universitat Polit\`ecnica de Catalunya, E-08034 Barcelona, Spain}
\author{G. Morigi}
\affiliation{Theoretische Physik, Universit\"at des Saarlandes, Campus E2.6, D-66123 Saarbr\"ucken, Germany}
\author{J. Boronat}
\affiliation{Departament de F\'isica, Campus Nord B4-B5, 
Universitat Polit\`ecnica de Catalunya,
E-08034 Barcelona, Spain}

\date{ March 21, 2024}

\begin{abstract}
We predict that ultracold bosonic dipolar gases, confined within a multilayer geometry, may undergo self-assembling processes, leading to the formation of chain gases and solids. These dipolar chains, with dipoles aligned across different layers, emerge at low densities and resemble phases observed in liquid crystals, such as nematic and smectic phases. We calculate the phase diagram using quantum Monte Carlo methods, introducing a newly devised trial wave function designed for describing the chain gas, where dipoles from different layers form chains without in-plane long-range order. We find gas, solid, and chain phases, along with quantum phase transitions between these states. Specifically, we predict the existence of quantum phase transitions from gaseous to self-ordered phases, as the interlayer distance is decreased. Remarkably, in the self-organized phases, the mean interparticle distance can significantly exceed the characteristic length of the interaction potential, yielding solids and chain gases with densities several orders of magnitude lower than those of conventional quantum solids.
\end{abstract}

\maketitle

\textit{Introduction.}
Self-assembled systems in ultracold atoms represent a fascinating and rapidly evolving area of research at the intersection of atomic physics, quantum optics, and condensed matter physics. 
Among the intriguing systems in this context are dipolar chain gases and solids, in which atoms with large magnetic or electric dipole moments interact with each other via long-range and anisotropic dipolar interactions, leading to the spontaneous formation of ordered structures. 
These self-assembled systems can exhibit a variety of exotic quantum effects, including dipolar quantum droplets~\cite{PhysRevLett.117.205301, PhysRevA.94.033619, PhysRevA.94.021602, PhysRevLett.122.130405, PhysRevResearch.1.033088,Kadau2016,Schmitt2016,Ferrier2016,Chomaz2016}, roton excitations in dipolar condensates~\cite{PhysRevLett.90.110402,PhysRevLett.90.250403,Chomaz2018}, and dissociation of a dipolar chain crystal~\cite{you2014many}.

Recent experiments with ultracold atomic gases have achieved significant advances in the observation of self-organized systems, such as superfluid stripes reported in platforms with spin-orbit interactions~\cite{Li2017Stripe,PhysRevLett.130.156001}.
Additionally, solids with superfluid properties (supersolids) have been observed in setups with long-range cavity-mediated interactions~\cite{Leonard2017Supersolid}. In these latter setups, although spatial translational symmetry is broken, the crystal periodicity is externally imposed by cavities. This limitation was recently surpassed in experiments with dipolar quantum gases, which reported a spontaneous formation of crystals of droplets~\cite{PhysRevX.9.021012,Bottcher2019Transient,PhysRevLett.122.130405}. The realization of an atomic quantum self-assembled system, in which continuous translational symmetry is naturally broken, remains an open challenge within the community of ultradilute gases.

\begin{figure}
\centering
\includegraphics[width=0.47\textwidth]{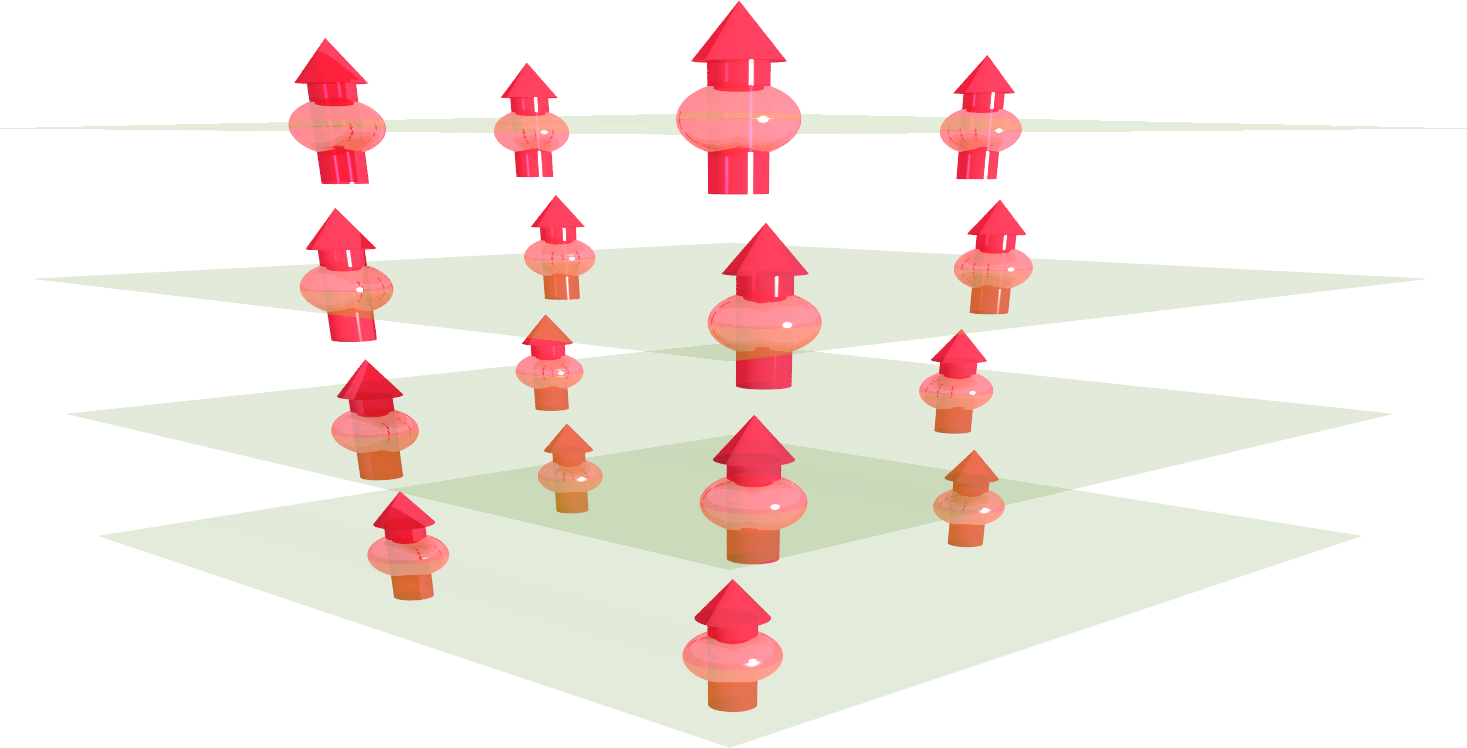}
\caption{
Schematic view of dipoles confined to several two-dimensional layers. The dipoles are assumed to be aligned perpendicularly to the planes by an external field. In the dipolar chain phase, there exists an ordering across the planes, as dipoles from different layers form chains, 
without in-plane long-range order.
}
\label{Fig:Setup}
\end{figure}

A relevant question that arises is if stable quantum solids and chains can be spontaneously formed at ultradilute densities, where the mean interparticle distance is much larger than the range of the potential. This scenario is unusual in condensed matter, where solid phases typically occur at high densities  
(a notable exception is that of Wigner crystals~\cite{PhysRev.46.1002} formed at low density; however, quantum effects are typically small). 
While there have been studies on low-density dipolar chains, previous research either neglected interactions between chains~\cite{PhysRevLett.97.180413} or restricted themselves to the regime of small spatial fluctuations, about the crystal phase~\cite{you2014many}. 
As a result, a complete description of self-assembled systems at ultradilute densities is still missing.

In the present article, we propose a possible physical realization of quantum self-assembled states. The platform consists of a system of dipolar bosons confined to a multilayer geometry, as schematically shown in Fig.~\ref{Fig:Setup}. Using exact many-body quantum Monte Carlo techniques~\cite{hammond1994monte,toulouse2015introduction,BoronatCasulleras1994} and proposing a novel trial wave function tailored specifically to model chain gases, we determine the ground-state phase diagram when the dipoles are all aligned perpendicularly to the parallel layers by an external field. We anticipate the emergence of a stable phase characterized by self-organized chains across layers when the density is low. We find a quantum phase transition from a gas to an ordered phase, solid or chains, as the interlayer distance decreases. Remarkably, the density of the self-assembled phases is several orders of magnitude lower than the one of conventional quantum solids.

\begin{figure}[!t]
\centering
\includegraphics[width=0.48\textwidth]{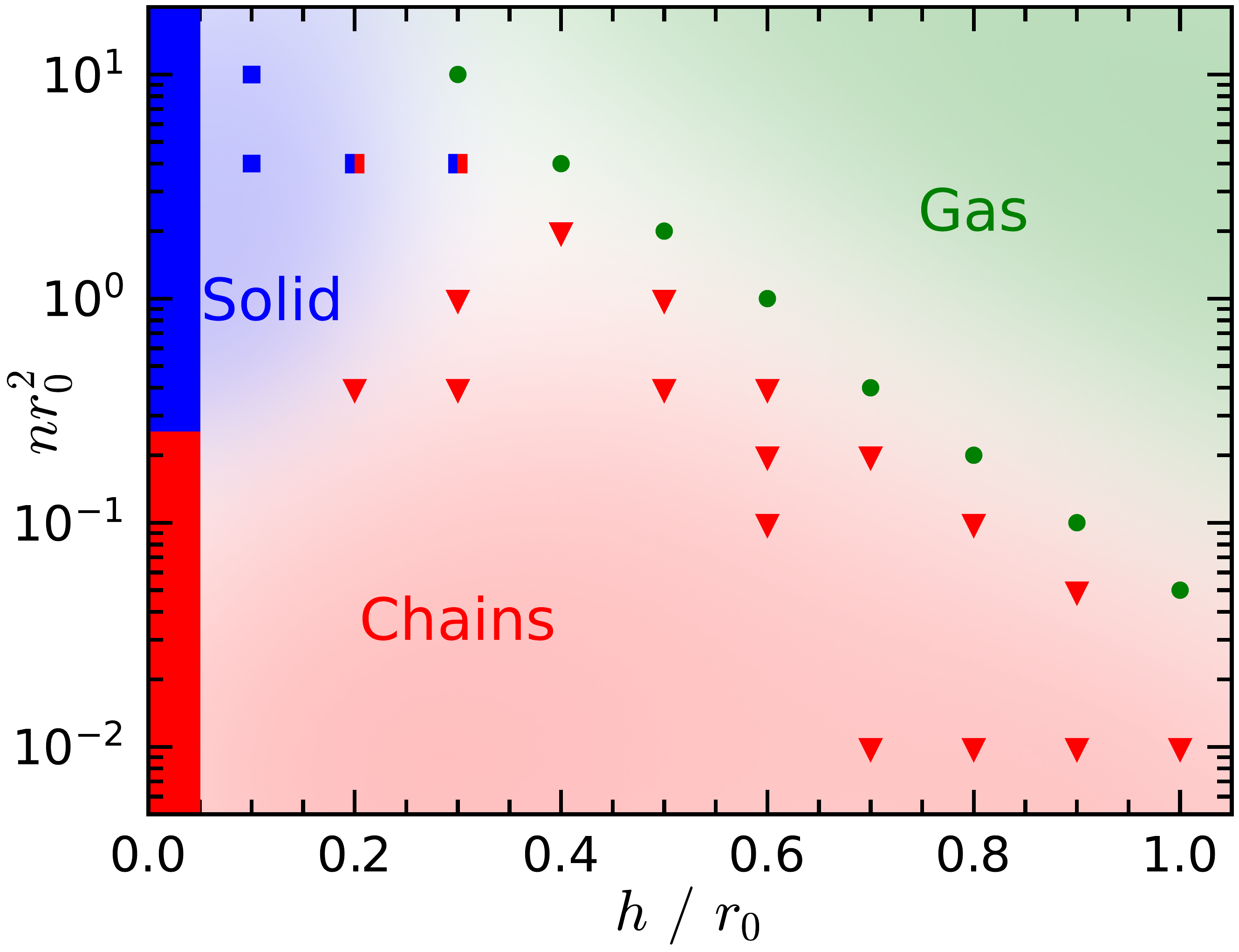}
\caption{
Ground-state phase diagram of dipolar bosons in a four-layer system at zero temperature as a function of the total density $nr_0^2$ and the interlayer distance $h/r_0$.  
Symbols indicate the phases according to the DMC calculations: triangles for chains, squares for solid, and dots for gas. 
Where solid and chain phases overlap, two-colored squares are used.
The coloring qualitatively shows the extension of each phase. 
As $h/r_0\to 0$,  
the low-density chain phase (red bar) transitions to a solid state (blue bar) at the critical density $n_{\rm c}r_0^2\approx0.283$.}
\label{Fig:PhaseDigram}
\end{figure}

\textit{Hamiltonian.} 
We consider $N$ bosons of mass $m$ and dipole moment $d$ constrained by a multilayer geometry composed of $M$ two-dimensional (2D) parallel layers, each separated by a distance $h$. We assume that each layer contains the same number ($N/M$) of dipoles and that the interlayer tunneling is suppressed so that the transverse degrees of freedom are frozen. The dipolar moments are considered to be oriented perpendicularly to the 2D planes, which can be achieved by applying a polarizing field. The Hamiltonian of this system is given by the sum of the kinetic energy operator and the dipolar interaction potential:
\begin{equation}
H\!=\!-\frac{\hbar^2}{2m}\sum_{i=1}^{N}\!{\bm{\nabla}}^2_i + d^2\!\sum_{i<j}^N\!\frac{|{\bf r}_i-{\bf r}_j|^2-2(l_i-l_j)^2h^2}{(|{\bf r}_i\!-\!{\bf r}_j|^2\!+\!(l_i\!-\!l_j)^2h^2)^{5/2}}.
\label{Hamiltonian}
\end{equation}
Here, ${\bf r}_i$ denotes $(x,y)$ coordinates of particle $i$ in one of $M$ planes with $l_i=1,2,\ldots,M$ being the index of the plane. For dipoles in the same layer, $l_i=l_j$, the dipolar interaction is always repulsive and falls off with a power law $1/r^3$. For dipoles in different layers, $l_i \neq l_j$, the dipolar interaction is attractive for small values of $r$ but repulsive for large values of $r$, where $r$ is the in-plane distance between dipoles. In the following, we use the dipolar length $r_0=md^2/\hbar^2$ as a unit of length. A further characteristic length is the interlayer distance $h$: A dipole at the origin of the layer $l_i=1$ experiences an attractive interaction with atoms at the $l_j$ layer within a cone of radius equal to $\sqrt{2}|l_i-l_j|h$.

\textit{Quantum Monte Carlo simulations.}
The ground-state properties of the system, described by the Hamiltonian~(\ref{Hamiltonian}), are calculated using the diffusion Monte Carlo (DMC) method~\cite{hammond1994monte,toulouse2015introduction,BoronatCasulleras1994}. This method has been shown to give an accurate description of correlated quantum systems~\cite{Ceperley}. The DMC algorithm solves stochastically the many-body Schrödinger equation in imaginary time resulting in an exact value (within controllable statistical error) of the ground-state energy, alongside other properties. As usual in DMC, we use guiding wave functions for importance sampling to reduce the variance. We use three trial wave functions: for a gas, solid, and chains. The positions of the dipoles are projected into a box of size $L_x\times L_y$ with periodic boundary conditions along $(x,y)$. The box size is related to the total density according to $n=N/(L_xL_y)$ (further details in the end matter section).
 
\textit{Ground state of dipolar bosons within a four-layer system.}
To construct the phase diagram, we explored the parameter space (dimensionless density $nr_0^2$ and interlayer distance $h/r_0$) and calculated the ground-state energy using the three different trial wave functions. The phase at each point corresponds to the phase that yields the lowest energy. We calculate the DMC energy $E_{\rm DMC}$, add the tail energy correction $E_{\rm tail}$, and extrapolate the corrected value to the thermodynamic limit, $E_{\rm th}$, as discussed in Supplemental Material
\footnote{See Supplemental Material, which includes Refs.~\cite{BoronatCasulleras1994,hammond1994monte,toulouse2015introduction,Ceperley,PhysRevA.99.023618,PhysRevLett.122.105302,PhysRevA.99.043609,PhysRevA.101.041602,PhysRevLett.128.063401,PhysRevResearch.2.023405,PhysRevA.100.023608,PhysRevLett.119.250402,PhysRevA.103.013311,PhysRevA.98.053632,PhysRev.155.88,RevModPhys.89.035003,phdthesisGG,Astrakharchik2007}, for additional information about the trial wave functions, freezing density, and finite size effects.}.
The resulting phase diagram is reported in Fig.~\ref{Fig:PhaseDigram} as a function of two key parameters: $nr_0^2$ and $h/r_0$. An important observation is the existence of the chain phase which spans across a wide range of interlayer distances. Its large extension enhances the feasibility of its experimental observation. Furthermore, this phase extends to extremely low densities, at which the mean interparticle distance within one layer can range from 1 to 20 times larger than the dipolar length $r_0$, resulting in the formation of atomic dipolar chains at ultradilute density. 
For $h/r_0\lesssim 0.2$, the strong interlayer attraction between dipoles results in the formation of tightly bound chains of a small $(x-y)$ spatial size.
As the density is increased, the strong in-plane dipolar repulsion pushes the system toward crystallization, which eventually occurs at the critical density $n_{\rm c}r_0^2\approx0.283$. 
While for small $h$, the radius of the attractive cone is small, resulting in rigid chains, for larger $h$ the chains become flexible and loosely bound. Consequently, an increase in the density for large values of $h$ leads to a transition from chains to gas rather than to a solid.
Remarkably, for certain densities ($nr_0^2\approx 1$), the decrease in $h/r_0$ induces a series of phase transitions crossing gas, chain, and solid phases. We provide estimates for the boundaries of the gas phase, along with the transition line from chains to gas. However, determining the precise transition from chains to solid is computationally challenging due to the similar energies of these phases across a broad parameter range.

\begin{figure*}[htp]
  \centering
  \subfigure{\includegraphics[width=0.32\textwidth]{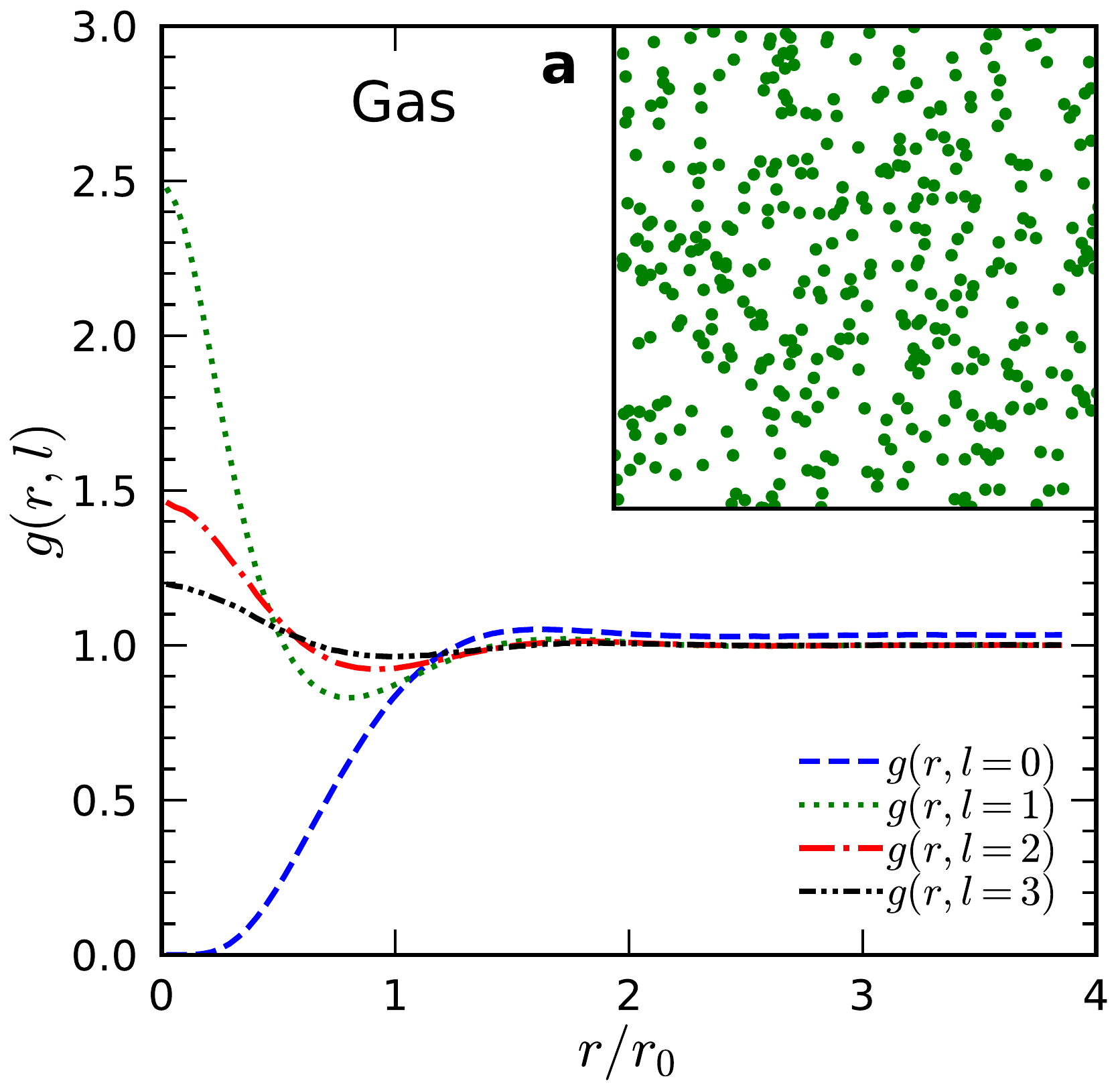}}
  \subfigure{\includegraphics[width=0.32\textwidth]{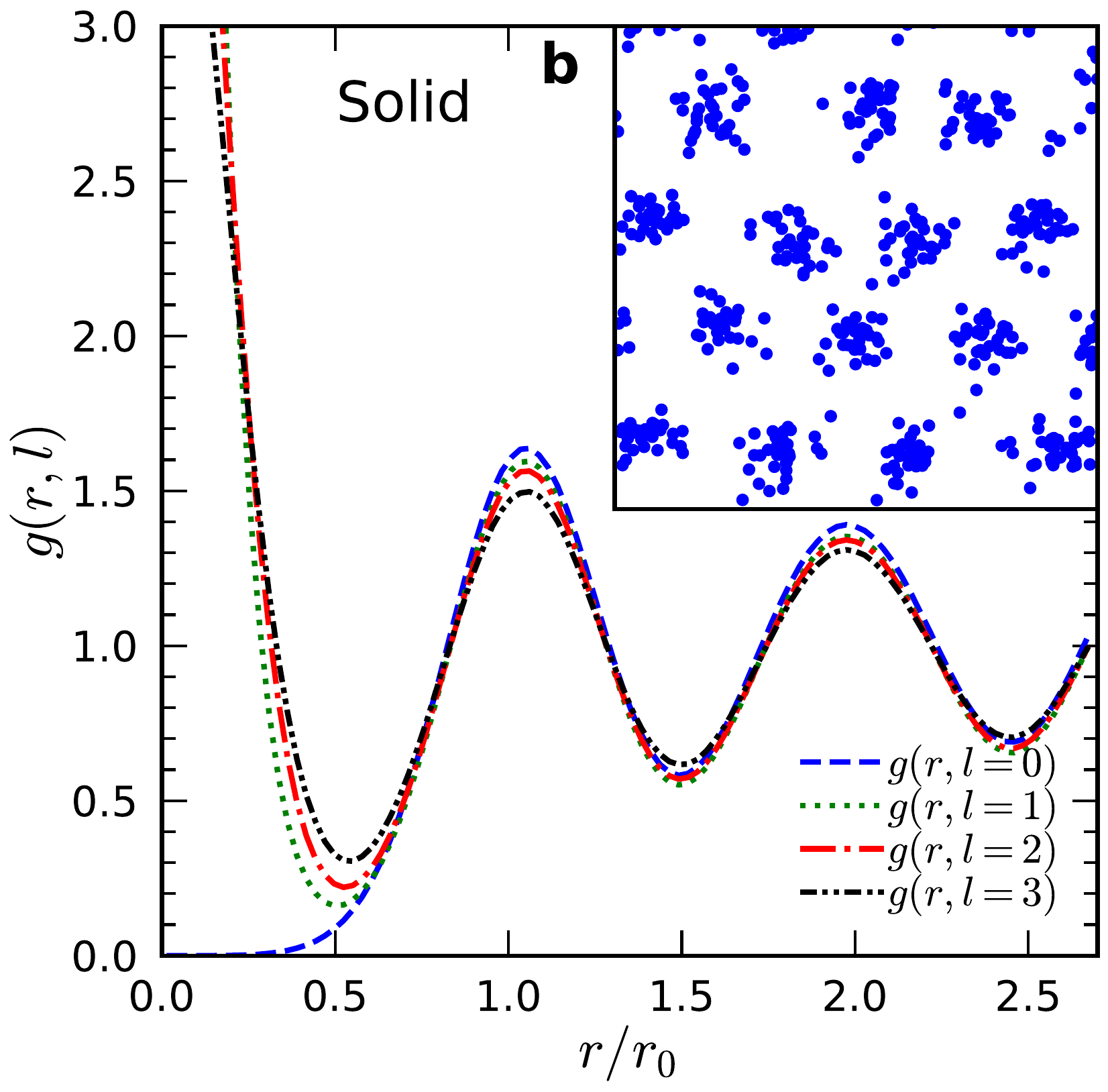}}
  \subfigure{\includegraphics[width=0.32\textwidth]{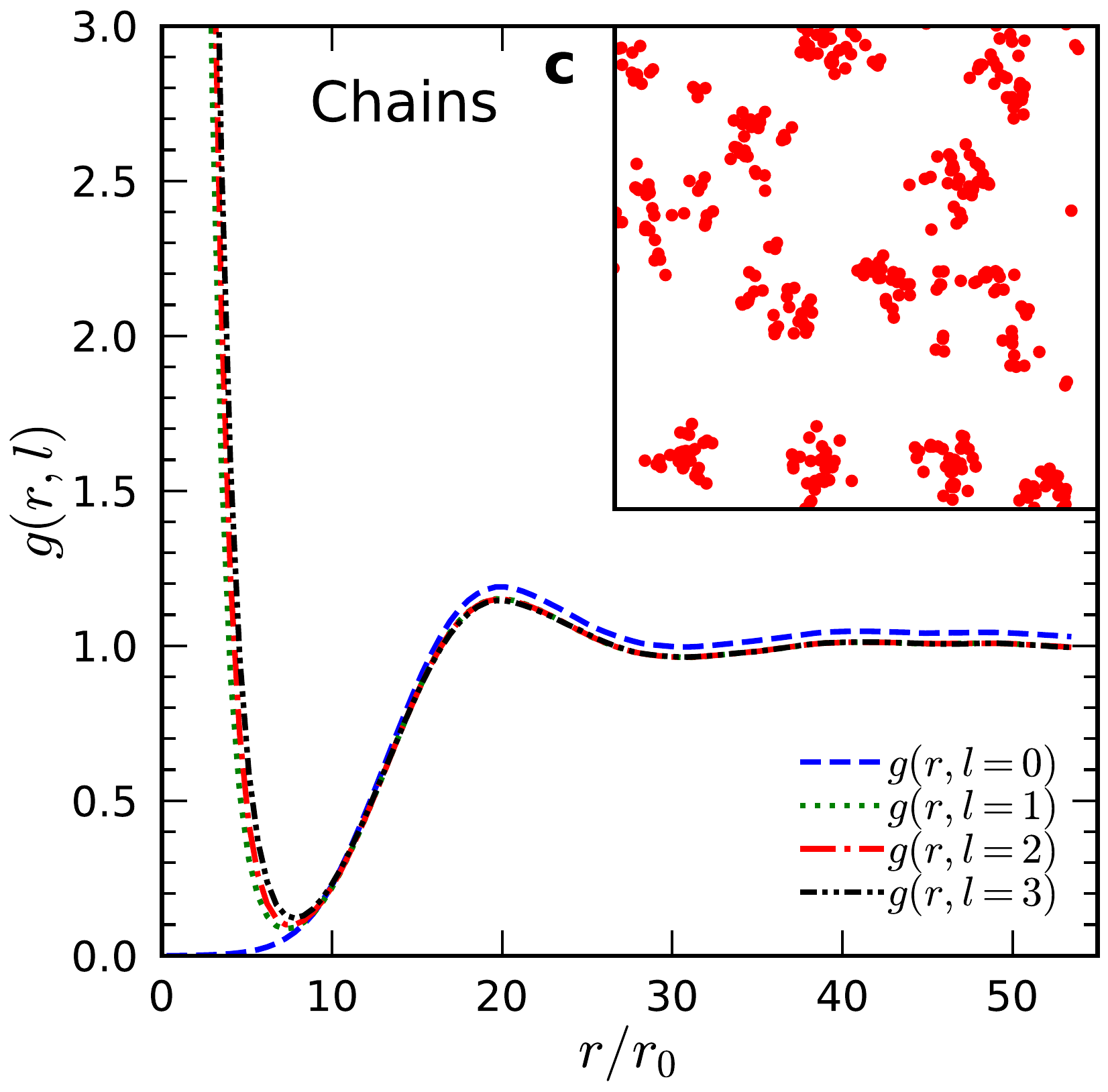}}
  \caption{Typical in-plane $g(r,l=0)$ and off-plane $g(r,l>0)$ pair distribution functions in the gas~(a), solid~(b), and chain~(c) phases within a four-layer geometry.
  Inset: characteristic snapshots of the atom coordinates in the corresponding phase.
The parameters are: gas, $nr_0^2=2.0$, $h/r_0=0.5$; solid, 
      $nr_0^2=4.0$, $h/r_0=0.3$; and chain, $nr_0^2=0.01$,  $h/r_0=0.9$.
      }  
     \label{Fig:PairDistributions}
 \end{figure*}

Various phases exhibit significantly different spatial distribution of dipoles, as quantified by the pair distribution function $g(r,l)$, which indicates the probability of finding two atoms at a relative in-plane distance $r$ and separated by $l$ layers. In Fig.~\ref{Fig:PairDistributions}, we show characteristic examples of the pair distribution functions $g(r,l)$ in the gas~(a), solid~(b), and chain~(c) phases.

In the gas phase, the in-layer distribution ($l=0$) vanishes at short distances due to a strong dipolar repulsion for $r\to 0$. For larger distances, $g(r,l=0)$ exhibits a shallow maximum and then approaches the unit value, i.e., the asymptotic value of uncorrelated atoms; see Fig.~\ref{Fig:PairDistributions}a. 
However, correlation functions between different layers ($l>0$) show a notably distinct short-range behavior. In particular, there is a high peak in $g(r,l>0)$ due to strong dipolar attraction for $r\to 0$.
For larger $r$, $g(r,l>0)$ exhibits a minimum, and eventually approaches its uncorrelated value of $1$. A representative snapshot of the atom positions in the gas phase in the DMC simulations is presented in the inset of Fig.~\ref{Fig:PairDistributions}a, showing that the dipoles are uniformly distributed without any specific pattern. 

Characteristic examples of the pair distribution functions in the solid phase are shown in Fig.~\ref{Fig:PairDistributions}b. At short distances, $r\to 0$, the behavior is similar to that observed in the gas phase, where the same-layer distribution vanishes as dipoles approach each other, while different-layer distributions exhibit a pronounced correlation peak. For larger distances, the pair distributions show stronger correlations in the solid phase as compared to the gas one, as a result of the spatial ordering in the solid. Consequently, $g(r,l)$ displays several pronounced oscillations in a solid phase, whereas in the gas phase, it rapidly converges to the uncorrelated values of $g(r,l)=1$. The amplitude of these oscillations depends on the system parameters. In a triangular lattice, the first peak is particularly prominent, corresponding to six neighbors located at the lattice spacing $r\approx a_\text{latt}$, which is proportional to the mean interparticle distance $a_\text{latt} = (2/\sqrt{3})^{1/2} r_m \approx 1.075 r_m$ and is determined by the in-layer density, $r_m = \sqrt{M/n}$. In the inset of Fig.~\ref{Fig:PairDistributions}b, a snapshot of the atom coordinates during the DMC simulation of the solid phase is presented, revealing the periodic structure of a triangular lattice. 

In Fig.~\ref{Fig:PairDistributions}c, we report the pair distribution functions $g(r,l)$ for the chain phase. We observe intermediate features, where some aspects resemble those of a gas phase while others are closer to those of a crystal phase. Specifically, the in-plane pair distribution $g(r,l=0)$ shows correlations typical of the gas phase, indicating no long-range order within the layer. However, the off-plane pair distributions $g(r,l>0)$ exhibit high peaks at short distances, similar to the strong interlayer correlations observed in the solid phase. The snapshots of the atom coordinates during the DMC simulation, shown in the inset in Fig.~\ref{Fig:PairDistributions}c, provide insight into the structure of this unconventional phase. Here, the dipolar chains are uniformly distributed without forming a regular structure, indicating the absence of positional order. However, the dipoles forming the chains exhibit strong correlations in the direction perpendicular to the planes. This behavior resembles that of a nematic phase in liquid crystals.

\begin{figure}[!t]
\centering
\includegraphics[width=0.48\textwidth]{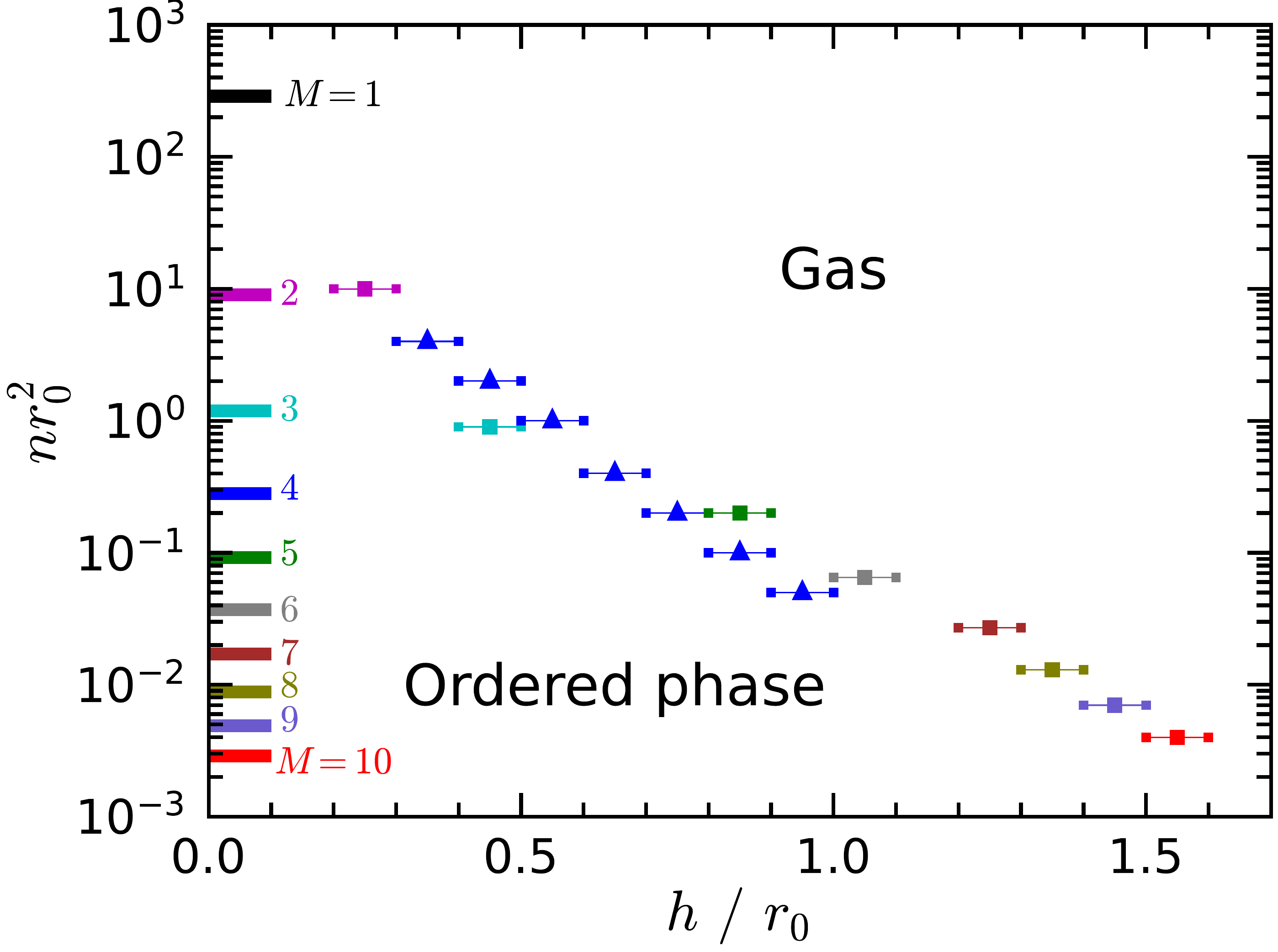}
\caption{Phase diagram for layers $M=1,\cdots,10$ as a function of the total density $nr_0^2$ and the interlayer distance $h/r_0$. A quantum phase transition from a gas to a self-ordered phase occurs as $h/r_0$ decreases.
Squares indicate DMC predictions for the gas-ordered transitions at a single value of $nr_0^2$ while varying $h/r_0$ for each $M$. 
Triangles denote DMC results for $M=4$, varying both $nr_0^2$ and $h/r_0$.
Thick lines show the crystallization densities $n_{\rm c}r_0^2$ for various $M$.}
\label{Fig:PhaseTransition}
\end{figure}

\textit{Quantum phase transition: From gas to a self-ordered phase.}
The effect of different numbers of layers (from one to ten) on the phase diagram is summarized in Fig.~\ref{Fig:PhaseTransition}, where the schematic phase diagram of the multilayer system is shown as a function of the dimensionless density $nr_0^2$ and the separation between layers $h/r_0$. The thick horizontal lines for small $h/r_0$ correspond to the critical freezing density $n_{\rm c}r_0^2=290/M^5$, indicating the density at which an $M$-layer system of dipolar bosons will form a crystal of superdipoles (where all dipoles align perfectly at each point) when the interlayer separations are extremely small $h\to0$ (see Supplemental Material~\cite{Note1}). Notably, this estimation of the freezing point shows that the freezing density decreases as $1/M^5$ with the number of planes $M$. This strong dependence on $M$ suggests that the multilayer geometry is a promising setup for observing quantum solids with low surface densities in future experiments. Increasing the number of planes can yield a very dilute solid, especially for small interlayer distances.

Using the DMC method, we observed a quantum phase transition from a gas to an ordered phase (solid or chains) as the interlayer distance $h/r_0$ decreased. To determine the critical interlayer values for this transition, we fixed the number of planes $M$ and the total density $nr_0^2$, and then systematically reduced the interlayer distance $h/r_0$ while keeping the density constant. We calculated the ground-state energy using both gas and ordered (solid or chains) trial wave functions. The phase at each point corresponds to the phase with the lowest energy. The critical interlayer distance corresponds to the value at which the phase changes from gas to an ordered phase. The estimated critical interlayer distances for the gas-ordered transition are reported in the phase diagram in Fig.~\ref{Fig:PhaseTransition}, covering the range from $M=2$ up to 10. For each $M$ value, the critical interlayer distance is reported for a single density, except in the case of $M=4$, where it is reported for multiple density values. Remarkably, the self-ordered phase exists across a broad range of densities and interlayer separations. 
Triangles indicate the gas-ordered phase transition for $M=4$ layers. At a fixed density, decreasing $h/r_0$ leads to a transition from gas to a self-ordered phase. This reduction in $h/r_0$ also diminishes the radius of the attractive cone of the dipolar interaction, aligning dipoles along the layers. Conversely, at a fixed $h/r_0$, decreasing the density induces to the transition. This is because lower density reduces in-plane repulsion, while interlayer attraction depends primarily on $h/r_0$. Overall, the transition density decreases as $h/r_0$ increases, and this behavior in the four-layer system is likely scalable to systems with more layers. Adjusting $M$ can significantly alter the density at which an ordered phase appears, decreasing from $nr_0^2=10$ to $0.001$ as $M$ increases from 2 to 10. Such extremely low densities correspond to mean interparticle distances ranging from $0.4$ to 100 times the dipolar length $r_0$. Moreover, the range of $h$ for stable self-assembled phases expands with $M$, which suggests the possibility of its experimental realization. 

We propose that the $M$-dipole condensate, defined as the macroscopic eigenvalue of the $M$-body density matrix, could be serve as an order parameter to distinguish between different phases. 
The atomic condensate ($M=1$) is finite in the gas phase but vanishes in the chain and solid phases. 
In the bilayer case, it has been shown that the molecular condensate ($M=2$) becomes finite when chains of length $M=2$ are formed~\cite{Macia2014}. Similarly, we expect that when a chain of length $M$ is formed, the $M$-dipole condensate is present while all other condensates formed from $M-1, M-2,\cdots, 1$ dipoles vanish. 
In contrast, condensation would be entirely absent in the solid phase.

\textit{Feasible experimental parameters.}
Recently, an experimental realization of a stack of head-to-head dysprosium atoms in bilayer arrays with interlayer distances in the sub-50-nm scale has been achieved at density $nr_0^2\approx 10^{-2}$~\cite{Du2023AtomicPO}. 
To form coherent chains, as in our DMC calculations, the thermal de Broglie length should be larger than the chain length.
This requires low enough temperatures, $k_BT \lesssim \hbar^2/(m(M-1)^2h^2)$, so that longer chains require lower temperatures. For the bilayer case, the temperature should be $T\lesssim\SI{1.2}{\micro\kelvin}$, which is close to experimental conditions ($T=\SI{1.7}{\micro\kelvin}$ in Ref.~\cite{Du2023AtomicPO}). 
An ultradilute quantum self-assembled dipolar system can be achieved using magnetic dipolar atoms or molecules such as $^{162}$Dy~\cite{Du2023AtomicPO}, $^{164}$Dy$_2$~\cite{PhysRevLett.107.190401}, and $^{168}$Er$^{164}$Dy~\cite{PhysRevLett.121.213601}, which have dipolar lengths of approximately $r_0\sim \SI{21}{\nm}$, $r_0\sim \SI{168}{\nm}$, and $r_0\sim\SI{123}{\nm}$, respectively 
(or $h/r_0 = 2.4; 0.3; 0.4$ assuming $h=\SI{50}{\nm}$). 
While the experiment~\cite{Du2023AtomicPO} considers head-to-head orientation, head-to-tail orientation can also be experimentally investigated.
Reaching lower temperatures could lead to strong interlayer correlations requiring a description beyond mean-field theory, such as our DMC calculations. 

\textit{Conclusions.}
We used the diffusion Monte Carlo method to obtain the phase diagram of two-dimensional dipolar bosons confined within a multilayer geometry at zero temperature. Our findings indicate the possibility of generating self-assembled phases, such as solid and chains, at ultralow densities. Remarkably, the density of these systems can be several orders of magnitude lower than that of conventional solids. Moreover, the freezing density can be further decreased by increasing the number of layers. Our results present a novel example of quantum dipolar self-ordered phases characterized by ultradilute density. In the solid phase, unlike previous realizations, the dipolar interactions by themselves are sufficient to create the crystalline structure, which consists of individual atoms rather than clusters. Whether this crystal exhibits a finite superfluid signal and qualifies as a new supersolid remains a topic for future research.

\begin{acknowledgments}
We gratefully acknowledge the fruitful discussions with Joaquim Casulleras.
We acknowledge financial support from Ministerio de Ciencia e Innovación MCIN/AEI/10.13039/501100011033 (Spain) under Grant No. PID2020-113565GB-C21 and from AGAUR-Generalitat de Catalunya Grant No. 2021-SGR-01411. GG and GM acknowledge partial support by the Deutsche Forschungsgemeinschaft (DFG, German Research Foundation), with the CRC-TRR 306 “QuCoLiMa”, Project-ID No. 429529648. GG acknowledges funding from the European Union-Next Generation, the Spanish Ministry of Universities and the Recovery, Transformation and Resilience Plan through a grant from Universitat Polit\`ecnica de Catalunya.
\end{acknowledgments}

\textit{End matter on trial wave functions.}
We use guiding wave functions to impose a specific symmetry on the state. 
We describe the gas phase by the Bijl-Jastrow (BJ) wave function,
\begin{equation}
\Psi_{\rm G}(\mathbf{r}_1,\dots,\mathbf{r}_N)=\prod_{i<j}^N
f_{2}(|\mathbf{r}_{i}-\mathbf{r}_{j}|),
\label{TWFGas}
\end{equation}
where $f_2(r)$ is a two-body correlation function.
The solid is described by a Nosanow-Jastrow (NJ) wave function,
\begin{equation}
\Psi_{\rm S}(\mathbf{r}_1,\dots,\mathbf{r}_N;\{\mathbf{R}_I^c\})=
\prod_{j<k}^N
f_{2}(|\mathbf{r}_{j}-\mathbf{r}_{k}|)
\prod_{i=1}^N
f_{1}(|\mathbf{r}_{i}-\mathbf{R}^c_{i}|),
\label{NosanowTWF}
\end{equation}
which contains one- and two-body terms $f_1(r)$ and $f_2(r)$, respectively. 
The key difference is that the BJ function maintains translational symmetry, while the NJ describes a state with a broken translational symmetry.

In addition to the gas and solid phases, the presence of anisotropic interactions allows for the emergence of a dipolar chain phase. This phase is similar to the molecular phase observed in bilayer systems~\cite{Macia2014,PhysRevA.94.063630,Cinti2017,PhysRevLett.128.063401} but extends beyond two layers. While the use of BJ (for gas) and NJ (for crystal) trial functions is well established in the literature, to the best of our knowledge, a Monte Carlo description for dipolar chain systems is still lacking.
Here, we propose a new type of trial wave function specifically designed to describe chain phases.
Its primary characteristic lies in not only capturing interlayer correlations but also enabling the free movement of chains within planes. 
In our model, a single chain consists of $M$ dipoles, each belonging to a different layer without any crystal ordering between them. The new trial wave function is constructed as follows:
\begin{equation}
\Psi_{\rm C}(\mathbf{r}_1,\dots,\mathbf{r}_N)=\prod_{j<k}^N f_{2}(|\mathbf{r}_{j}-\mathbf{r}_{k}|)
\prod_{i=1}^N f_{M}(|\mathbf{r}_{i}-\mathbf{R}_{i}|).
\label{TWFChains}
\end{equation}
where $f_{M}(|\mathbf{r}_{i}-\mathbf{R}_{i}|)$ is a many-body term, depending on the position of an atom $\mathbf{r}_i$ and the center of mass of the chain $\mathbf{R}_i=~(\sum_{k\in{\rm C}_i}\mathbf{r}_k)/M$, with the sum covering all atoms belonging to the same chain as the $i$th atom.
We consider $f_{M}(r)$ to have a Gaussian shape given by $f_{M}(r_i)=e^{-\alpha|\mathbf{r}_{i}-\mathbf{R}_{i}|^2}$, where the variational parameter $\alpha$ quantifies the strength of localization.  
The function $f_{M}(r)$ is a many-body term, as a displacement of a single atom leads to $M$ changes in the function due to the resulting movement of the center of mass.
A crucial difference between the solid and chain trial wave functions is that the positions of the lattice sites $\mathbf{R}_i$ in the NJ function remain static and are fixed by the triangular lattice, whereas they are allowed to change in $\Psi_{\rm C}$ during the simulation.
We found that the chain trial wave function results in lower energy in certain regions of the phase diagram compared to the trial wave functions for solids and gases. Therefore, the novel introduction of the chain trial wave function is essential for fully describing the phase diagram of the dipolar multilayer system (further details can be found in the Supplemental Material~\cite{Note1}).

\section{Supplemental Material}
\section{Diffusion Monte Carlo Method}
The ground-state properties of the system 
are calculated using the diffusion Monte Carlo (DMC) method (for details of the method see, for example, Refs.~\cite{BoronatCasulleras1994,hammond1994monte,toulouse2015introduction}). This method has demonstrated its capacity to provide an accurate description of correlated quantum systems~\cite{Ceperley}. Examples include ultradilute  bosonic~\cite{PhysRevA.99.023618,PhysRevLett.122.105302} and fermionic mixtures~\cite{PhysRevA.99.043609}, few and many-bosonic bound-states~\cite{PhysRevA.101.041602,PhysRevLett.128.063401}, Bose~\cite{PhysRevResearch.2.023405} and Fermi~\cite{PhysRevA.100.023608} polarons, dipolar Bose supersolid stripes~\cite{PhysRevLett.119.250402}, Bose gas subject to a multi-rod lattice~\cite{PhysRevA.103.013311}, and ultracold quantum gases with spin-orbit interactions~\cite{PhysRevA.98.053632}.
The DMC method solves the many-body Schrödinger equation in imaginary time, yielding a precise estimation of the ground-state energy within controllable statistical error. By using guiding wave functions, it is possible to impose a certain symmetry characteristic for a given phase and study phase transitions. Also, the use of guiding wave functions acts as in importance sampling, meaning that the more relevant regions of phase space are better sampled, minimizing the variance and the statistical error.
To describe the gas phase we chose a translationally invariant trial wave function in Bijl-Jastrow (BJ) form, composed as a pair product of two-body correlation terms, while for the solid phase we used a Nosanow-Jastrow (NJ) wave function, incorporating one- and two-body correlations factors and breaking the translational symmetry. To describe the chain phase we introduce a new trial wave function, combining two-body correlation terms with many-body terms.

\section{Trial wave function: Gas phase}
To describe the gas phase we chose a trial wave function of the Bijl-Jastrow form, composed as a pair product of two-body correlations terms
\begin{equation}
    \Psi_{\rm G}(\mathbf{r}_1,\dots,\mathbf{r}_N)=\prod_{i<j}^N
f_{2}(|\mathbf{r}_{i}-\mathbf{r}_{j}|).
 \label{TWFGas}
\end{equation}
The two-body terms $f_2(r)$ depend on the distance between a pair of atoms $r_{ij}=|\mathbf{r_i}-\mathbf{r_j}|$.
Here, the two-body terms $f_{2}(r)$ are taken as the solution of the two-body problem at short distances. This solution depends on whether the dipoles are in the same layer or not. 
For dipoles in the same layer, we chose the short-distance part of the two-body correlations term as
\begin{equation}
    f_{2}(r)=C_1 K_0\left(2\sqrt{R_0r_0/r}\right),
\end{equation}
up to $R_{\rm match}$. Here $K_0(r)$ is the modified Bessel function, $R_{\rm match}$ and $R_0$ are variational parameters.

For distances larger than $R_{\rm match}$ we chose the two-body correlation function as
\begin{equation}
f_{2}(r)=C_2 {\rm exp}\left[ -\frac{C_3}{r} + \frac{C_3}{r-L}\right],
\end{equation}
which take into account the contributions from other particles and describe long-range phonons in the form established by Reatto and Chester~\cite{PhysRev.155.88}. The coefficients $C_1$, $C_2$, and $C_3$ are fixed by imposing continuity of the function and its first derivative at the matching distance $R_{\rm match}$, and also that $f_{2}(L/2)=1$.

For dipoles in different layers, the interlayer correlations are taken as the solution of the two-body problem $f_{2}(r)$ up to $R_1$. We also impose the boundary condition $f^{'}_{2}(R_1)=0$. For distances larger than the variational parameter $0<R_1<L/2$ we set $f_{2}(r)=1$.

\section{Trial Wave Function: Solid phase}
We use the standard Nosanow-Jastrow trial wave function to describe the solid phase~\cite{RevModPhys.89.035003}
\begin{equation}
    \Psi_{\rm S}(\mathbf{r}_1,\dots,\mathbf{r}_N;\{\mathbf{R}_I^c\})=
\prod_{j<k}^N
f_{2}(|\mathbf{r}_{j}-\mathbf{r}_{k}|)
\prod_{i=1}^N
f_{1}(|\mathbf{r}_{i}-\mathbf{R}^c_{i}|),
\label{NosanowTWF}
\end{equation}
where $\mathbf{r}_i$ are the positions of the atoms, $\{\mathbf{R}_I^c\}$ are the positions defining the equilibrium crystal lattice, and $f_1(r)$ and $f_2(r)$ are the one-body and two-body correlation factors, respectively. 

The two-body correlation functions $f_2(r)$ used in $\Psi_{\rm S}$ are of the same form as those used to describe the gas phase $\Psi_{\rm G}$, although the specific values of the variational parameters might differ.

The one-body terms $f_1(r)$ localizes each atom close to its lattice site and are modeled by a Gaussian function
\begin{equation}
    f_1(r_i)=e^{-\alpha|\mathbf{r}_{i}-\mathbf{R}^c_{i}|^2},
\end{equation}
where $\alpha$ is the localization strength and $R_i^c$ is the
position of the lattice site. Notice that in our case, $R_i^c$ are fixed by the
triangular lattice. Meanwhile, $\alpha$ is a variational parameter, which
is chosen by minimizing the variational energy.

\section{Trial wave function: Chain Gas phase}
A single chain consists of $M$ dipoles, each belonging to a different layer without any crystal ordering between them. These chains are flexible but will not become tangled due to repulsive forces between dipoles in the same layer. To describe the chain phase, we construct a trial wave function by combining two-body correlation terms $f_2(r)$, which depend on the distance between a pair of atoms $r_{jk}=|\mathbf{r}_j-\mathbf{r}_k|$, with many-body terms $f_{M}(|\mathbf{r}_{i}-\mathbf{R}_{i}|)$, depending on the position of an atom $\mathbf{r}_i$ and the center of mass of the chain $\mathbf{R}_i$ (which depends on the positions of atoms in the same chain).
\begin{equation}
\Psi_{\rm C}(\mathbf{r}_1,\dots,\mathbf{r}_N)=\prod_{j<k}^N f_{2}(|\mathbf{r}_{j}-\mathbf{r}_{k}|)
\prod_{i=1}^N f_{M}(|\mathbf{r}_{i}-\mathbf{R}_{i}|).
\label{TWFChains}
\end{equation}
For the two-body functions $f_2(r)$ we maintain the same structure as in the gas and solid phases, although the specific values of the variational parameters are generally different. We consider the many-body terms $f_{M}(r)$ of a Gaussian shape, 
\begin{equation}
f_{M}(r_i)=e^{-\alpha|\mathbf{r}_{i}-\mathbf{R}_{i}|^2},
\end{equation}
where the variational parameter $\alpha$ quantifies the strength of localization. The chain center of mass is defined as 
\begin{equation}
\mathbf{R}_i=\frac{1}{M}\sum_{k\in{\rm C}_i}\mathbf{r}_k,
\end{equation}
with the sum covering all atoms belonging to the same chain as the $i$-th atom. The function $f_{M}(r)$ is a many-body term, as a displacement of a single atom leads to $M$ changes in the function due to the resulting movement of the center of mass.
A crucial difference between the solid and chain trial wave functions is that the positions of the lattice sites $\mathbf{R}_i$ in the NJ function remain static and are fixed by the triangular lattice whereas they change in $\Psi_{\rm C}$ during the simulation.
We found that using the chain trial wave function $\Psi_{\rm C}$ lowers the energy in specific parts of the phase diagram compared to the trial wave functions used for solids and gases. This means that introducing the chain trial wave function is crucial for accurately describing the phase diagram of the dipolar multilayer system (further details of this trial wave function can be found in Ref.~\cite{phdthesisGG}).

\section{Freezing density}
In Ref.~\cite{Astrakharchik2007}, the authors studied the ground-state phase diagram
of a two-dimensional Bose system with dipole-dipole interactions using the QMC methods. The dipoles were constrained to move in a single plane and were polarized in the perpendicular direction to the plane. The authors found a quantum phase transition from a gas to a solid phase as the density increases.
This transition was estimated to occur at the critical density $\tilde{n}\tilde{r}_0^2\approx290$,
with $\tilde{r}_0=r_0=md^2/\hbar^2$ and $\tilde{n}=n$ for a single layer of dipoles. 

Now, we show how $\tilde{n}\tilde{r}_0^2\approx290$ is
rewritten for a system with $M$ layers in the limit of rigid chains. To do this,
consider a chain with $M$ dipoles, one in each layer. 
When $h\to0$ the chain becomes a super-dipole with
mass $Mm$, dipolar moment $Md$, and dipolar length
\begin{equation}
    \tilde{r}_0=\frac{(Mm)(Md)^2}{\hbar^2}=
    M^3r_0.
    \label{Eq:Sdipoler0}
\end{equation}
Now, consider an $M$-layer system with $N$ dipoles and $N/M$ chains evenly distributed.
When $h\to0$ the $M$-layer system effectively becomes to a single-layer one, each chain becomes a super-dipole,
and the number of particles changes from $N$ dipoles to $N/M$ super-dipoles.
As a consequence of the latter the density now is given by $\tilde{n}=n/M$, and in dipolar units it becomes
\begin{equation}
    \tilde{n}\tilde{r}_0^2=
    \left(\frac{n}{M}\right)\left(M^3r_0\right)^2
    =M^5nr_0^2.
\end{equation}
From the last relation and with $\tilde{n}\tilde{r}_0^2\approx290$ we obtain
\begin{equation}
    n_cr_0^2=\frac{290}{M^5}.
\end{equation}
In the limit of $h\to0$ an $M$-layer system of dipolar bosons will
crystallize at the critical density $290/M^5$.

\section{Finite size effects}
\begin{figure}[h]
	\centering
    \subfigure{\includegraphics[width=0.47\textwidth]{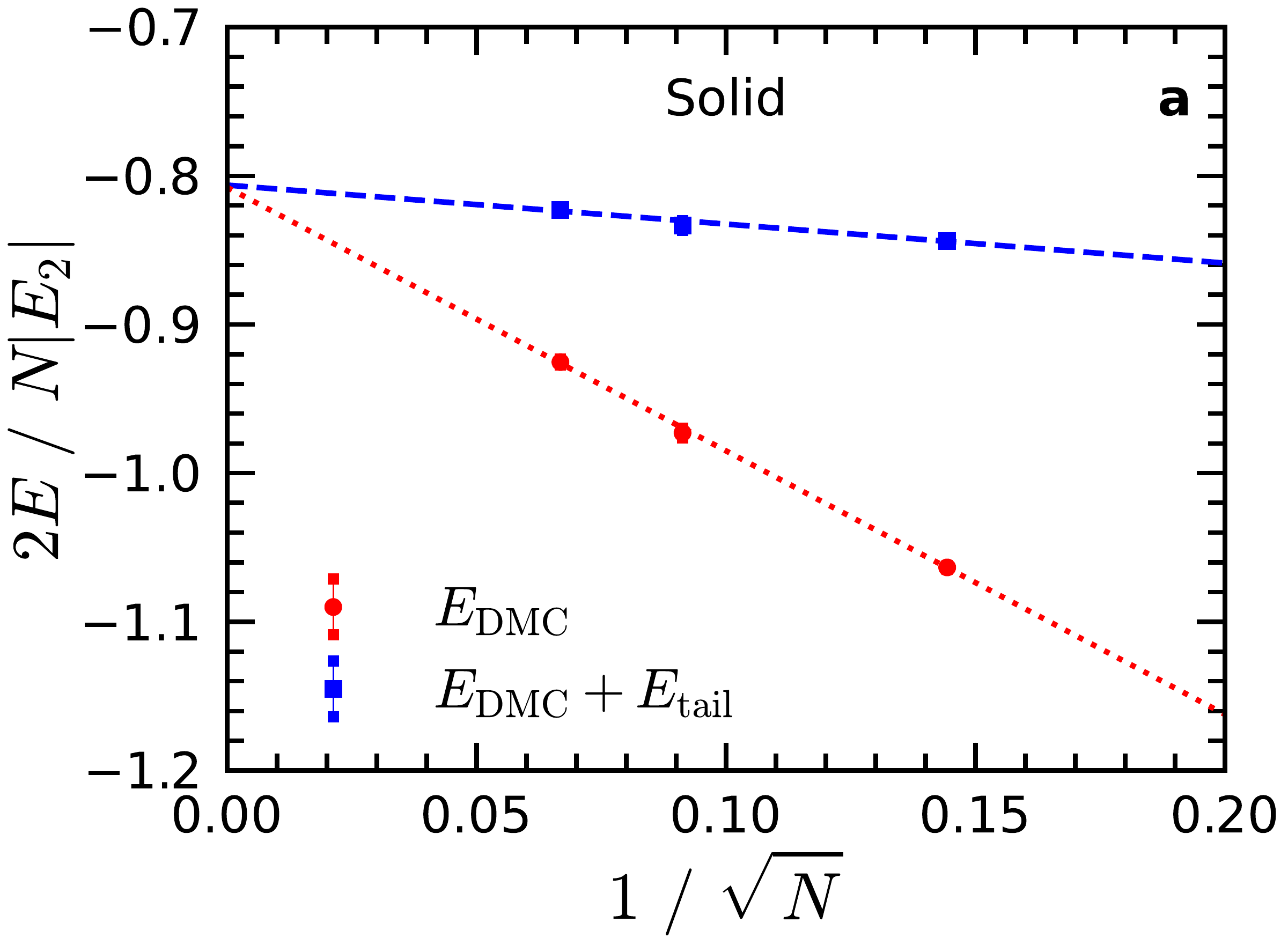}}
    \subfigure{\includegraphics[width=0.48\textwidth]{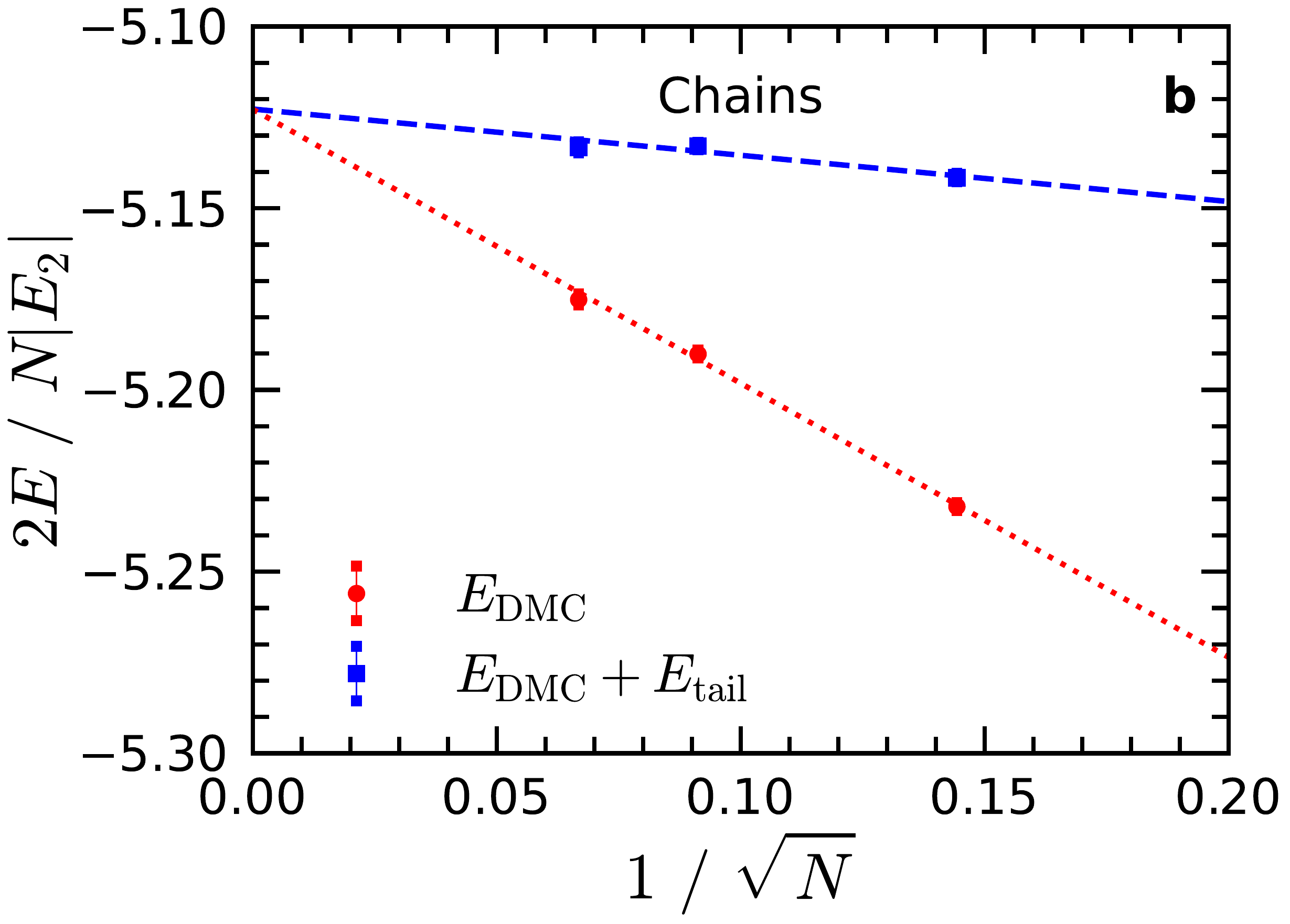}}
    \caption{Examples of the finite-size dependence for the energy
        in the solid (a) and chain (b) phases. Solid phase parameters: $nr^2_0=0.4$, $M=4$, and $h/r_0=0.5$. Chain phase parameters: $nr^2_0=0.01$, $M=4$, and $h/r_0=0.9$.
        Symbols, DMC energy and DMC energy with added the tail energy;
        curves, fit $E_{\rm th}+C/\sqrt{N}$.}
\label{Fig:FiniteSize}
\end{figure}

In two dimensions, the dipolar potential is a quasi-long ranged one, therefore its truncation at $L/2$ produces significant finite-size corrections. 
The average energy $E_{int}$ of the interaction potential $V(r)$ which is a two-body operator, can be expressed in terms of the pair distribution function $g(r)$ as
\begin{equation}
    \frac{E_{int}}{N} = \frac{1}{2n}\int_0^\infty V(r) g(r) d{\bf r}.
\end{equation}
Applied to the multilayer geometry, the finite-size effects can be significantly diminished by adding the tail energy,
\begin{equation}
\begin{aligned}
\frac{E_{{\rm tail}}(n,L)}{L^2}=\int_{L/2}^{\infty}& \bigg[ 
\frac{M}{2}\frac{d^2}{r^3}g_{\sigma\sigma}(r)+
\frac{M(M-1)}{2}\frac{d^2(r^2-2h^2)}{(r^2+h^2)^{5/2}}g_{\sigma\sigma'}(r) \bigg]2\pi rdr .
\label{eq5}
\end{aligned}
\end{equation}
The intraspecies and interspecies pair distribution functions are denoted by $g_{\sigma\sigma}(r)$ and $g_{\sigma\sigma'}(r)$, respectively. An approximate value of the tail energy~(\ref{eq5}) is obtained by ignoring correlations at large distances, $g_{\sigma\sigma}(r)\to n_{\sigma}^2$ and $g_{\sigma\sigma'}(r)\to n_{\sigma}n_{\sigma'}$, which leads to
\begin{equation}
\frac{E_{{\rm tail}}}{N}=\frac{2\pi d^2 n^{3/2}}{M\sqrt{N}}+
\frac{2\pi d^2(M-1)N}{M(4h^2+N/n)^{3/2}}.
\label{eq6}
\end{equation}
In Fig.~\ref{Fig:FiniteSize}, we show two examples of the finite-size study for the energy. 
In it, we consider solid and chain phases.
Notice that the addition of the tail energy~(\ref{eq6}) to the DMC data allows to significantly reduce the finite-size dependence. 
After adding the tail energy $E_{\rm tail}$ to the DMC energy $E_{\rm DMC}$,
we extrapolate the energy $E(N)=E_{\rm DMC}+E_{\rm tail}$ to the thermodynamic limit
value $E_{\rm th}$ using the fitting formula
\begin{equation}
    E(N)=E_{\rm th}+\frac{C}{\sqrt{N}}, 
\label{Eq:6.10}
\end{equation}
where C is a fitting parameter.
For both phases we observe that the energy dependence on the number of particles scales as
$1/\sqrt{N}$, contrary to the law $1/N$, usual for fast decaying potentials. 
We find that our fitting function describes
well the finite-size dependence. 

The number of particles used in this study ranges from $N=48$ up to $N=224$. All the results of the energy reported in our work are corrected to the thermodynamic limit using this functional law. The same procedure is repeated for all the system parameters for the gas, chains, and solid phases.
 
\begin{figure}[h]
	\centering
    \includegraphics[width=0.45\textwidth]{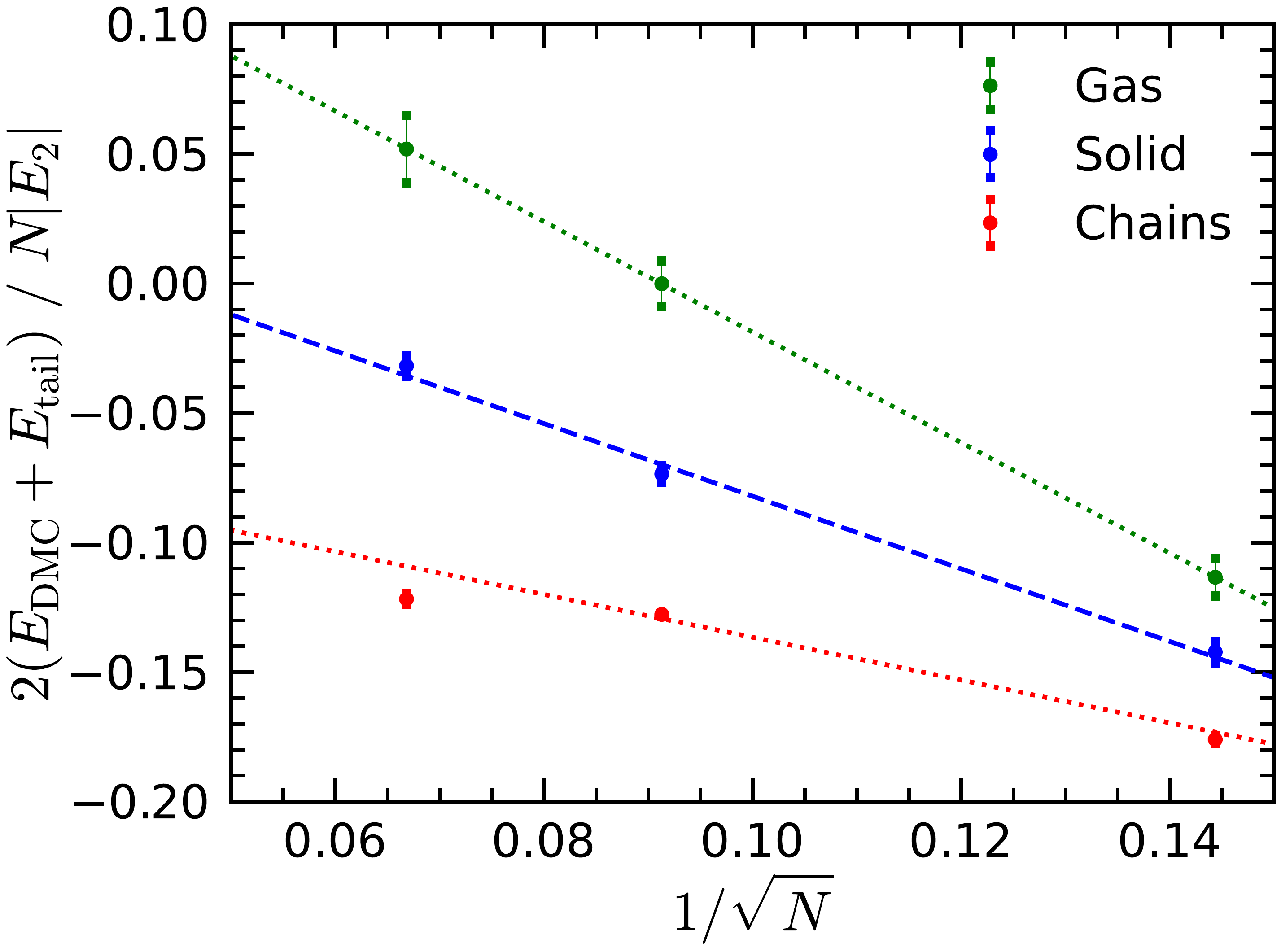}
    \caption{Example of the chain phase being energetically favorable compared to the gas and solid phases.
        Phase parameters: $M=4$, $h/r_0=0.6$ y $nr_0^2=0.4$.
        Symbols, DMC energy with added the tail energy;
        curves, fit $E_{\rm th}+C/\sqrt{N}$.}
\label{Fig:ComparationEnergy}
\end{figure}

In Fig.~\ref{Fig:ComparationEnergy} we show the energies of a gas, solid and chains for $M=4$, $h/r_0=0.6$ y $nr_0^2=0.4$.
The energies are extrapolated to the thermodynamic limit by using the empirical law $E(N)=E_{th}+C/N^{1/2}$, with $C$ a fitting parameter, and we 
have added the tail energy to reduce the finite-size effects.  
For the particular values shown in the figure, the chain phase is energetically favorable compared to the gas and solid phases.

%
    
\end{document}